# Modeling and Optimization of Two-Terminal Spin-Orbit-Torque MRAM

Md Nahid Haque Shazon, Piyush Kumar, *Graduate Student Member, IEEE,* Luqiao Liu, Daniel C. Ralph, and Azad Naeemi, *Senior Member, IEEE*

*Abstract*—This paper presents physical modeling and benchmarking for two-terminal spin-orbit torque magnetic random-access memory (2T-SOT-MRAM). The results indicate that the common SOT materials that provide only in-plane torque can provide little to no improvement over spin-transfer-torque (STT) MRAM in terms of write energy. However, emerging SOT materials that provide out-of-plane torques with efficiencies as small as 0.1 can result in significant improvements in the write energy for such 2-terminal devices, especially when the magnet lateral dimensions are scaled down to 30 or 20 nm. Additionally, a novel 2T-SOT MRAM device is proposed that can increase the path electrons pass through the SOT layer; hence, increasing the generated spin current and the energy efficiency of the device. Our benchmarking results indicate that an out-of-plane SOT efficiency of 0.051 for 20nm wide devices can result in write energies approaching SRAM at the 7nm technology node.

## I. INTRODUCTION

AS semiconductor technology progresses towards advanced technology nodes, nonvolatile memory technologies, such as resistive random-access memory (RRAM) [1], ferroelectric random-access memory (FeRAM) [2], spin transfer torque random-access memory (STT-MRAM) [3], and spin-orbit-torque magnetic random-access memory (SOT-MRAM) [4] are being pursued as potential options to meet the ever-growing demand for on-chip memory. Among these options, SOT-MRAM has risen as a viable option for future memory solutions. While FeRAM is potentially the most energy-efficient option thanks to its voltage-controlled write mechanism, it still faces major challenges such as large write voltages and its destructive read mechanism [2]. STT-MRAM (Fig. 1(a)) is now commercially available in older technology generations. However, its adoption at advanced technology nodes faces major challenges because of the large write currents that are required [4].

SOT-MRAM (Fig. 1(b)) utilizes the spin-orbit coupling effect to change the magnetization of the free layer ferromagnet of a magnetic tunnel junction (MTJ) via the spin Hall effect [5]. In contrast to STT-MRAM, which employs a current through the MTJ to change the magnetization [6], SOT-MRAM uses an in-plane current passing through a non-magnetic (typically a heavy metal) layer to produce a perpendicular spin current that exerts a torque on the magnetization of the free layer of the MTJ [7]. It is worthwhile noting that the spin-transfer torque per unit charge (in units of ℏ/2) is always <1 in STT-MRAM, whereas SOT-MRAM can have a spin-transfer torque per unit charge >1 depending on spin diffusion length and device geometry, making it more energy-efficient compared to STT-MRAM [8]. SOT-MRAM also demonstrates better read performance, improved endurance, and lower power consumption. These advantages make SOT-MRAM particularly appealing for applications requiring high-speed, energy-efficient, and reliable memory solutions, such as in-memory computing, probabilistic computing, and reservoir computing [9].

The traditional SOT-MRAM design, however, is a three-terminal (3T) structure, consisting of two transistors and one MTJ per cell [10]. Despite the benefits this scheme offers, the 3T configuration presents significant challenges in terms of cell size and memory density, especially at advanced nodes [11]. Another significant challenge in conventional perpendicular magnetic anisotropy (PMA) SOT-MRAM is that common high-symmetry spin source materials can only generate SOT with the torque direction in the sample plane. In the presence of a symmetry-breaking applied magnetic field, an in-plane spin torque can manipulate PMA layers in micron-scale devices via domain wall motion, but it is incapable of driving the form of anti-damping switching needed for efficient control of single-domain PMA layers on the scale of 30 nm and below [12]. Recent research has concentrated instead on materials that can generate spin polarization in the out-of-plane direction, which should enable efficient anti-damping switching of nanoscale perpendicular magnets [14-16]. Nonetheless, obtaining sufficiently high out-of-plane spin torque efficiency in practical materials continues to be a challenge, and it remains uncertain whether materials with the elevated out-of-plane values necessary for effective switching will become available. It is also worthwhile to mention that although the traditional SOT-MRAM design can beat STT-MRAM, its projected performance is still far from SRAM [7].

A 2-terminal (2T) SOT-MRAM (Fig. 1(c)) was proposed in [17] and later was experimentally demonstrated in [18], which reduces the number of transistors required per cell from two to one (making it a two-terminal device instead of three), thereby

Manuscript submitted 29 October 2025. This work was supported by SUPREME: Superior Energy-efficient Materials and Devices, one of seven centers in JUMP 2.0, an SRC program sponsored by DARPA.

Md Nahid Haque Shazon, Piyush Kumar, and Azad Naeemi are with the School of Electrical and Computer Engineering, Georgia Institute of Technology, Atlanta, GA 30332 USA (e-mail: mshazon3@gatech.edu).

Luqiao Liu is with the Department of Electrical Engineering and Computer Science, Massachusetts Institute of Technology, Cambridge, MA 02139, USA.

Daniel C. Ralph is with the Department of Physics and Kavli Institute at Cornell, Cornell University, Ithaca, NY, 14853.

significantly decreasing the overall cell size and increasing memory density. In addition, the 2T structure does not require an external magnetic field for reliable switching since the byproduct STT current can help break the symmetry, thus further simplifying the device structure and reducing fabrication complexity [19]. However, there is a notable lack of detailed investigations into the energy efficiency of the 2T scheme, with its reduced cell size and potentially altered magnetic dynamics as compared to the STT-MRAM. Similarly, while studies like [20] have explored material innovations for optimizing 3T-SOT MRAM, there remains a gap in understanding what kind of SOT materials would provide the lowest write energy for 2T-SOT MRAM. This highlights the need for targeted studies on material-structure co-optimization specifically for the 2T-SOT MRAM.

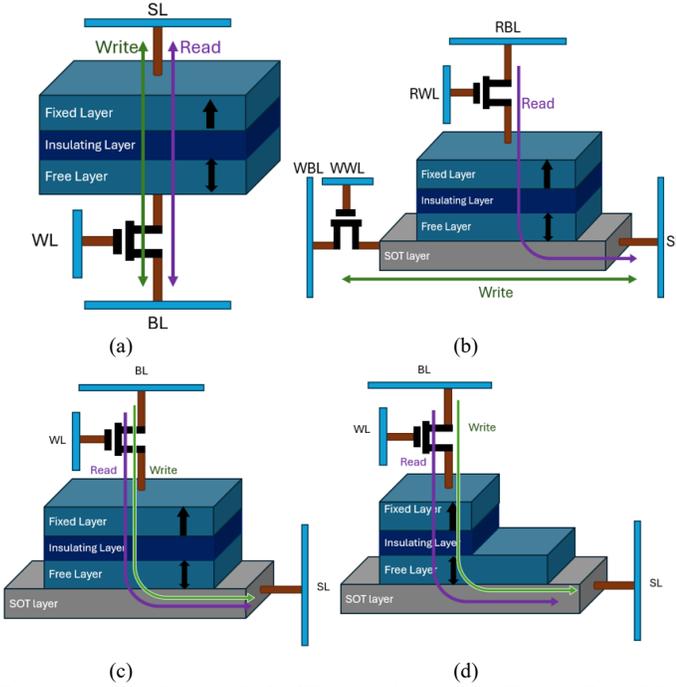

Fig. 1. (a) STT-MRAM cell, (b) 3T conventional SOT-MRAM cell, (c) 2T SOT-MRAM cell with full fixed layer, (d) 2T SOT-MRAM cell with half-fixed layer.

To address these gaps, we present a comprehensive material-device modeling, optimization, and benchmarking study for SOT-MRAM. We use a drift-diffusion modeling framework that uses the electric current distribution obtained from finite-element simulations and calculate the generated spin current. The current crowding in 2T-SOT MRAM shortens the path of most electrons through the SOT-layer. To mitigate this effect and improve the generated spin current, we propose a device with a half-fixed layer as shown in Fig. 1(d). However, for both full and half-fixed layer devices, we find that SOT layers with only in-plane SOT efficiency ($\xi_{DL,y}$) offer little improvement over STT-MRAM in terms of write energy, and an out-of-plane SOT component ($\xi_{DL,z}$) is needed to substantially lower the write energy and potentially reach SRAM-level performance. However, the $\xi_{DL,z}$ needed to approach the SRAM write energy is substantially lower than what is needed in the case of 3T-SOT.

The rest of the paper is organized as follows. Section II describes the modeling approach for calculating the generated spin current in various devices, assuming an SOT layer of AuPt. Section III delineates the approach used to calculate the switching time and write energy for the case of AuPt or any possible in-plane SOT efficiencies. Section IV explores the benefits provided by an out-of-plane SOT and the array-level benchmarking. Finally, the work is summarized and concluded in Section IV.

## II. SPIN CURRENT MODELING FRAMEWORK

The spin current consists of spin Hall drift and diffusion contributions and can be expressed as [20]:

$$\vec{j}_s^z(z) = -\frac{\sigma_{NM}}{2e}\frac{\partial \vec{\mu}_s}{\partial z} - \sigma_{SH} E_x \hat{y}, \qquad (2)$$

where $\sigma_{NM}$ is the conductivity of the NM layer, $\vec{\mu}_s$ is the spin accumulation, $E_x$ is the applied electric field, and $\sigma_{SH}$ is the spin Hall conductivity, which is related to the spin Hall angle $\theta_{SH}$. Finally, the distribution of the spin current density along the z-axis is given by,

$$\vec{j}_s(z) = -[-j_{s0}^{SH}(-t)\hat{y} + (j_{s0}^{SH}(0)\hat{y} + \vec{j}_s^F) \\ \times \sinh\left(\frac{z+t}{\lambda_{sd}}\right)]/\sinh\left(\frac{t}{\lambda_{sd}}\right) - \hat{y} j_{s0}^{SH}(z), \qquad (3)$$

where $j_{s0}^{SH}(z) = \theta_{SH} J_x(z) = \sigma_{SH} E_x(z)$ [22]. A simplified model has been used in most prior studies to calculate the conduction current distribution between the ferromagnetic and non-magnetic layers, using the current distribution between two lumped resistances [23]. However, this simple model is not accurate enough for nanoscale devices because of the incomplete redistribution of the conduction current between the FM and NM layers over a short distance [24]. Additionally, in the case of 2T-SOT, current crowding is dominant, and the electric current varies significantly along the length of the SOT layer. Hence, for all three cases, we utilize finite element simulations [25] to accurately calculate the conduction current inside the NM layer and employ the spin drift-diffusion approach to estimate the spin current profile for nanoscale devices. The spin current at the NM and FM interface (z=0) is plotted using (3) with the material and device parameters [24] provided in Table I.

TABLE I
MATERIAL AND DEVICE PARAMETERS FOR THE SPIN CURRENT PLOT

| Symbol | Parameter | Values |
|---|---|---|
| $\lambda$ (nm) | Spin diffusion length | 1.7 |
|  | NM dimensions (nm) | $300 \times 42 \times 4$ |
| $d_N$ (nm) | NM thickness | 4 |
| $\theta_{SH}$ | Spin Hall angle | 0.58 |
| $\rho_{NM}$ ($\mu\Omega - cm$) | NM layer resistivity | 83 |
|  | FM dimensions (nm) | $42 \times 42 \times 1.3$ |
| $\rho_{FM}$ ($\mu\Omega - cm$) | FM layer resistivity | 110 |

In our work, CoFeB serves as the FM material, while AuPt represents the SOT material for the resistive SOT layer. Note that AuPt can only provide an in-plane spin torque efficiency and will not be the SOT layer when out-of-plane spin torque efficiency is considered. The conduction current profiles and spin currents generated by both SOT and STT for the conventional 3T SOT-MRAM (with no built-in STT), 2T SOT-MRAM with full fixed layer, and 2T SOT-MRAM with half-fixed layer are displayed in Fig. 2 based on spin current calculation using (3). For all three cases, the total conduction current is 60 µA.

The spin current density profile depicted in Fig. 2(a) for the traditional 3T SOT-MRAM configuration appears to be almost constant ($8.5 \times 10^{10} A/m^2$) within the central region (ranging from around 5 nm to 37 nm), with a minor decrease at the boundaries. The reason behind such a spin current profile is explored in [26]. In the case of 2T SOT-MRAM with a full fixed layer shown in Fig. 2(b), there is an almost linear increase in the spin current along the x-direction, which denotes a more localized spin current towards one edge of the device. Note that although the maximum value of the spin current density (around $10 \times 10^{10} A/m^2$) is higher in this case, the average spin current is lower than that of the 3T conventional scheme when the same conduction current is used. Notably, the spin current due to STT is constant throughout the x-axis because of the large resistance of the oxide layer (Red dashed line in Fig. 2(b)).

For the 2T SOT-MRAM with a half-fixed layer, the spin current reaches its maximum value ($8.5 \times 10^{10} A/m^2$) at the midpoint of the free layer and remains almost constant thereafter, as shown in Fig. 2(c). Furthermore, the spin current due to STT is prevalent from 0 nm to 21 nm and is double the STT current in the full fixed layer case. Therefore, the 2T with a half-fixed layer scheme is likely to exhibit improved magnetization dynamics due to the higher average SOT current density. Note that in the case of PMA MTJs, spin polarizations of the SOT- and STT-generated spin currents are perpendicular.

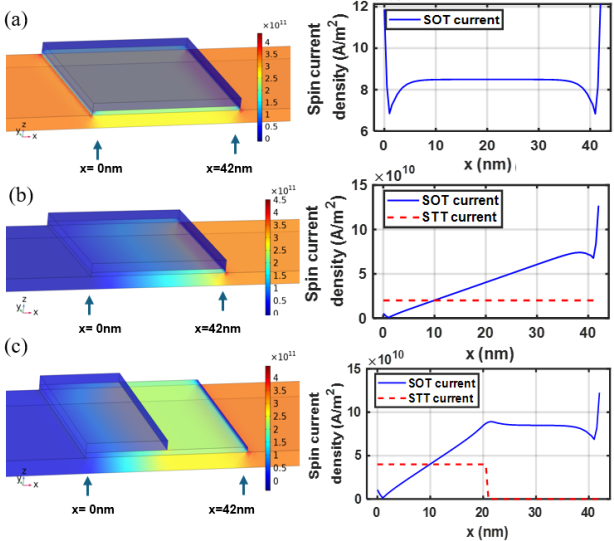

Fig. 2. COMSOL structures with conduction current density distribution and spin current density profile for (a) 3T conventional SOT-MRAM cell, (b) 2T SOT-MRAM cell with full fixed layer, (c) 2T SOT-MRAM cell with half-fixed layer.

## III. WRITE ENERGY FOR PURE IN-PLANE SOT

In this section, we compare the estimated write energies and switching probabilities of the two flavors of the 2T SOT-MRAM with those of the STT-MRAM and conventional 3T SOT-MRAM, through micromagnetic simulations. Note that only in-plane spin torque efficiency is considered in this section. The Object Oriented Micromagnetic Framework (OOMMF) [27] is utilized to conduct micromagnetic simulations where the simulation parameters listed in Table II are used. For STT-MRAM, only the AP-P scenario with a charge-to-spin conversion ratio of 0.6 is considered throughout this study [28]. Note that the P-AP case has a charge-to-spin conversion ratio of 0.3 and yields a lower spin current than the AP-P case. Thus, for comparing STT-MRAM performance with other MRAM configurations, only the best-case scenario of AP-P is taken into account. The modeling frameworks for STT and SOT that are used here are validated in [24] and [28]. Due to the computational overhead of using position-dependent spin current, we consider an error rate of less than 1% for all options. The Landau-Lifshitz-Gilbert-Slonczewski (LLGS) equation models how the magnetization in a material evolves over time under the influence of an applied spin torque [29]. The LLGS equation is expressed as:

$$\frac{dM}{dt} = -\gamma M \times H_{\text{eff}} + \frac{\alpha}{M_s}\left(M \times \frac{dM}{dt}\right) + \frac{\gamma}{\mu_0 M_s}\tau_{\text{SOT}}, \quad (1)$$

where M is the pointwise magnetization (A/m), $H_{\text{eff}}$ is the pointwise effective field (A/m), γ is the gyromagnetic ratio (m/(As)), α is the damping coefficient, and $\tau_{\text{SOT}}$ is the spin-orbit torque.

TABLE II
PARAMETERS FOR OOMMF SIMULATIONS

| Parameter | Values for PMA |
|---|---|
| Saturation magnetization $M_s$ (A/m) | $1.257 \times 10^6$ |
| Damping constant α | 0.01 |
| MTJ dimension (nm) | 42*42*1.3 |
| Resistance-area product ($\Omega \mu m^2$) | 8.5 |
| Magnetic anisotropy energy $K_u$ (MJ/$m^3$) | 1.27 |
| Temperature (K) | 300 |
| Thermal stability factor (Δ) | 57 |
| Exchange constant $A_x$ (pJ/m) | 20 |
| External magnetic field $B_x$ (mT) | 0 |

The write energy can be calculated using the following expression.

$$E_{write} = I_{cond}^2(R_{eff} + R_{MTJ} + R_O)t_{sw}, \quad (4)$$

where, $I_{cond}$ is the input conduction current, $R_{eff}$ is the equivalent resistance of the NM and FM layers in parallel, $R_{MTJ}$ is the MTJ resistance and $R_O$ represents the resistance of the access transistor and interconnects. The value of $R_O$ depends on the technology node and we use $R_O$=3.3kΩ assuming the 7nm CMOS technology node [24].

With AuPt as the SOT layer, the switching probability never approaches 100% for either of the 2T-SOT geometries no matter what current is applied or how long. This is because of the large in-plane spin torque efficiency of AuPt (~0.3) that results in the free magnet precessing near the horizontal plane as long as the write current is being applied. Once the current is turned off, the magnet may flip to either direction because of the thermal noise, and the write operation becomes indeterministic regardless of the write time. Hence, we look at a range of $\xi_{DL,y}$ values to study the impact of the SOT layer and identify the most promising value, as shown in Fig. 3(a). We also quantify the switching probability to ensure deterministic switching. The SOT-generated spin current accelerates the switching because of a shorter incubation time; hence, write energy decreases as $\xi_{DL,y}$ increases from $10^{-3}$ to $10^{-1}$. However, the improvement is small because of the weak SOT current. For $\xi_{DL,y}>0.1$, on the other hand, the stronger in-plane spin current results in a sudden drop in switching probability, as was explained for the case of AuPt. Fig. 3(b) shows a hypothetical case in which a symmetry-breaking field is provided for the 2T configurations. Results show that even with an external magnetic field, no meaningful benefits in terms of write energy can be achieved when only in-plane SOT is considered.

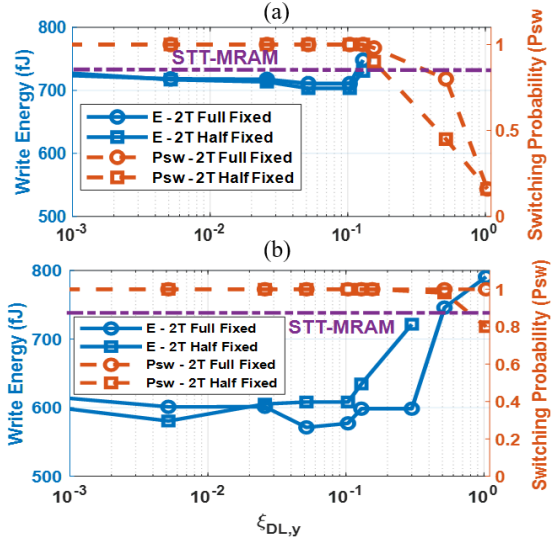

Fig. 3. Write energy vs in-plane ($\xi_{DL,y}$) spin torque efficiency (a) 2T case without Bx (b) 2T case with Bx

Notably, as seen from Fig. 3(a) and (b), the write energy values for 2T schemes are much higher than the SRAM write energy level for any $\xi_{DL,y}$, which is around 27 fJ. Note that the 3T conventional case is not shown in Fig. 3(a) because, without any symmetry breaking, no deterministic switching is noticed in this case. To reduce the write energy significantly and potentially approach the write energy of SRAM, an out-of-plane ($\xi_{DL,z}$) spin torque is needed; otherwise, the benefit offered by the 2T-SOT MRAM over STT-MRAM is going to be marginal and may not justify the added complexity. For the remainder of the paper, when $\xi_{DL,z}$ is present, the 3T SOT-MRAM, 2T SOT-MRAM with full and half-fixed layers will be referred to as 3T-ZSOT, 2T-ZSOT full FL, and 2T-ZSOT half FL, respectively.

## IV. BENEFITS OF OUT-OF-PLANE SOT EFFICIENCY AND DISCUSSIONS

Till now, we have investigated the impact of in-plane SOT efficiency. To study the role of the out-of-plane SOT, the write energies for various SOT-MRAM configurations against $\xi_{DL,z}$ for a constant $\xi_{DL,y}$ of 0.3 are plotted as shown in Fig. 4.

Notably, the 2T-ZSOT with full fixed layer and half-fixed layer demonstrate similar write energies while $\xi_{DL,z}$ is below 0.15. In addition, the write energy values for 2T schemes are found to be well below the STT level for all $\xi_{DL,z}$. For low $\xi_{DL,z}$ regions, 3T-ZSOT requires very high write energy since a large conduction current is required to provide sufficient spin current for magnetization reversal in the absence of any other symmetry-breaking mechanism. While the 3T-ZSOT scheme yields better write energy compared to the 2T-ZSOT configurations when the $\xi_{DL,z}$ is above 0.15, the 2T-ZSOT schemes demonstrate lower write energy when $\xi_{DL,z}$ is below 0.15. Fig. 4 demonstrates that the 3T-ZSOT scheme requires $\xi_{DL,z} = 0.052$ to reach the STT-MRAM write energy level. It is also noticed that to reach the SRAM write energy level, the 3T-ZSOT, 2T-ZSOT with full fixed layer, and 2T-ZSOT with half fixed layer require $\xi_{DL,z} = 0.39$, $0.67$, and $0.4$, respectively.

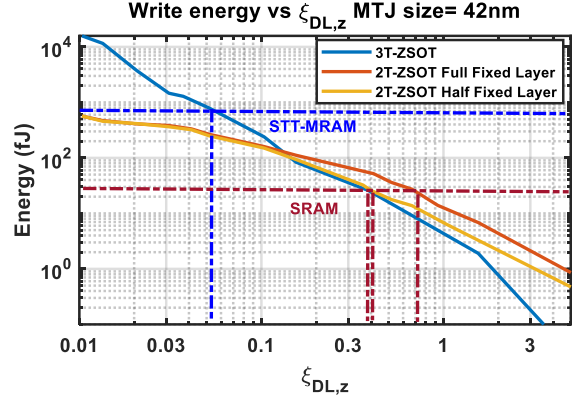

Fig. 4. Write energy vs out-of-plane ($\xi_z$) spin torque efficiency for MTJ size= 42nm.

Notably, STT efficiency increases for smaller devices since critical current scales with device area [30]. Also, when the total electric current is constant, a smaller magnet results in a larger current density in the SOT layer, which can compensate for the shorter path for spin current absorption. Hence, the next step is to investigate the impact of MTJ scaling on the required out-of-plane spin torque efficiency. Fig. 5 shows the values of $\xi_{DL,z}$ required to match the write energy levels of SRAM and STT-MRAM across various SOT-MRAM configurations as MTJ dimensions change. Note that the RA products for various MTJ sizes are modified accordingly to keep the MTJ resistances constant across those MTJ dimensions, which are adopted from [31]. The magnetic anisotropy is modified to keep the thermal stability constant across different MTJ sizes.

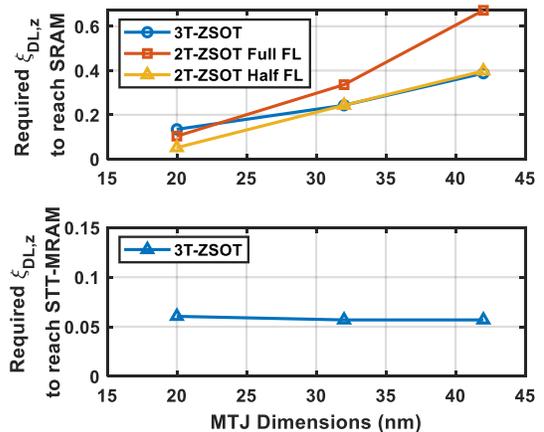

Fig. 5. Required out-of-plane spin torque efficiency ($\xi_{DL,z}$) to reach SRAM and STT-MRAM for different SOT-MRAM configurations.

As the size of the MTJ decreases, the necessary $\xi_{DL,z}$ to achieve SRAM-level write energy also decreases across all three configurations. Importantly, when the MTJ size becomes less than 32 nm, the 2T-ZSOT with half-fixed layer configuration starts to require a lower $\xi_{DL,z}$ compared to the 3T-ZSOT design to attain SRAM-level energy. Among all three configurations, the 3T-ZSOT architecture exhibits the least sensitivity to MTJ scaling in terms of the required $\xi_{DL,z}$, likely because it does not have any STT component, which can benefit from scaling. On the other hand, the 2T schemes gain more significantly from aggressive scaling since both SOT and STT current densities increase in this case. Since the 2T configurations virtually do not require any $\xi_{DL,z}$, only the $\xi_{DL,z}$ requirement for the 3T-ZSOT scheme to align with STT-MRAM write energy is presented in Fig. 5. It is worthwhile to mention that the results in Fig. 5 for the 3T-ZSOT are more demanding compared to prior work [20], which was based on 14nm technology node with smaller transistor and interconnect resistances ($R_o$ = ~2.5kΩ).

All the analyses so far are for a fixed $\xi_{DL,y}$ of 0.3, and for material exploration, it is important to obtain the design space in terms of both in-plane and out-of-plane SOT efficiencies, which are studied in Fig. 6 for all three device options, assuming a 20nm MTJ. Fig. 6(a) demonstrates that the 3T-ZSOT displays high write energy consumption, especially at lower spin torque efficiency. The write energy transitions to a relatively reduced state as $\xi_{DL,z}$ rises, suggesting that the increased spin torque efficiencies contribute to the decreased write energy. It is noticed that for reaching the write energy of STT-MRAM and SRAM, the 3T-ZSOT requires at least $\xi_{DL,z}$ = 0.06 and $\xi_z$ = 0.13, respectively. Notably, the write energy is mostly independent of the $\xi_y$ component. It is also noticed that the ratio $\xi_{DL,z}$ / $\xi_{DL,y}$ must exceed 0.16, 0.12, and 0.016 to guarantee dependable magnet switching at this dimension for 3T, 2T with full fixed layer, and 2T with half-fixed layer, respectively. Not conforming to this causes the white portion of Fig. 6(a), which indicates that no magnetization switching is achieved.

As shown in Fig. 6(b), the 2T-ZSOT with a full fixed layer shows reduced write energy when compared to the 3T configuration, particularly at low $\xi_{DL,y}$ and $\xi_{DL,z}$ regions. This improvement can be attributed to the inherent STT current present in the 2T configuration. It is noticed that for reaching the write energy of SRAM, the 2T-ZSOT with a full fixed layer requires at least $\xi_{DL,z}$ = 0.1. Notably, it virtually does not need any out-of-plane spin torque efficiency to reach the STT-MRAM level due to its built-in STT current.

Fig. 6(c) depicts the write energy for the 2T-ZSOT featuring a half-fixed layer, which demonstrates the lowest energy usage among the configurations assessed. The contour plot indicates that even at moderate $\xi_{DL,y}$ and $\xi_{DL,z}$ values, the write energy remains lower compared to other schemes. This enhancement can be attributed to the increased average spin current density and inherent STT current of this configuration. In addition, for reaching the write energy of SRAM, the 2T-ZSOT with a half-fixed layer requires around $\xi_{DL,z}$ = 0.051. Similar to its full fixed layer counterpart, it virtually does not need any out-of-plane spin torque to reach the STT-MRAM level. It is worthwhile noting that recently [32] has reported that TaIrTe$_4$ with a lower crystal symmetry can achieve a $\xi_{DL,z}$ of 0.11. The issue now is to find a material with substantial $\xi_{DL,z}$ that is manufacturable — but does not require epitaxial growth or single-crystal van der Waals flakes.

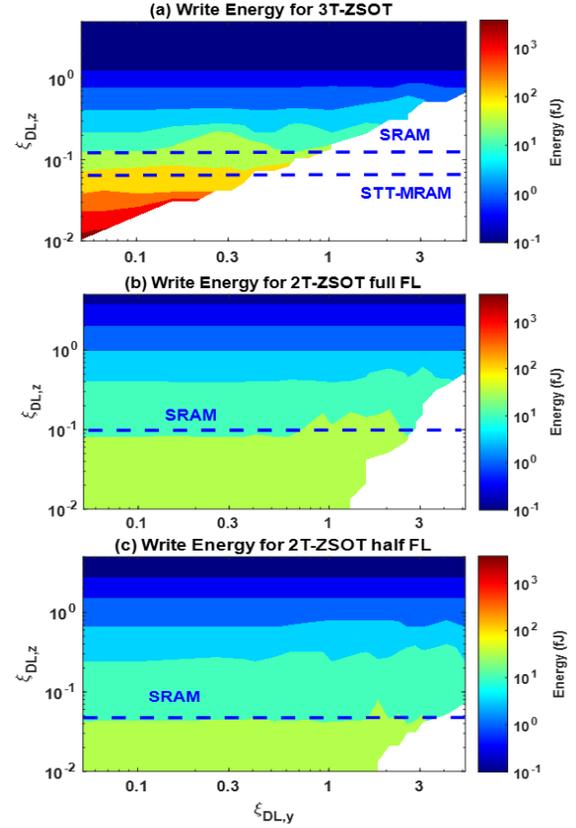

Fig. 6. The write energy contour plot with in-plane ($\xi_{DL,y}$) and out-of-plane ($\xi_{DL,z}$) spin torque efficiency for (a) 3T-ZSOT SOT-MRAM cell, (b) 2T-ZSOT SOT-MRAM with full fixed layer, (c) 2T-ZSOT SOT-MRAM with half fixed layer.

Since the write energy mostly depends on the out-of-plane spin torque efficiency, next, Fig. 7 shows the write energy for various SOT-MRAM configurations against only $\xi_{DL,z}$ for an MTJ size of 20nm. Notably, the 2T-ZSOT with the half-fixed layer scheme demonstrates lower write energy than any other scheme when $\xi_{DL,z}$ is lower than 0.2.

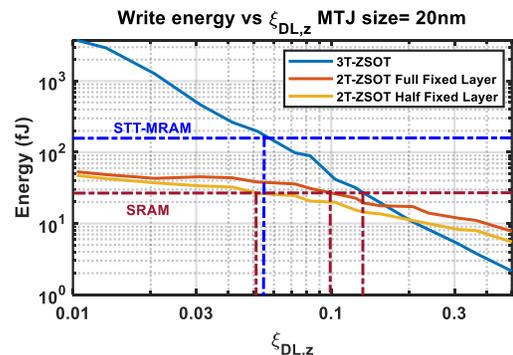

Fig. 7. Write energy vs out-of-plane ($\xi_{DL,z}$) spin torque efficiency for MTJ size = 20 nm.

## V. Conclusion

This study presents a comprehensive evaluation of two-terminal SOT-MRAM devices and demonstrates that, while in the absence of an out-of-plane spin torque efficiency ($\xi_{DL,z}$), they provide a marginal improvement over STT-MRAM, they can become far more competitive if a small $\xi_{DL,z}$ is present, and they become even more competitive as the MTJ dimensions are scaled down to ~20nm. To increase the path of the electric current inside the SOT layer and hence the generated spin current, an MTJ in which the fixed layer is half the size of the free layer is proposed. For a 20nm MTJ, while the conventional 3T-ZSOT scheme requires a minimum $\xi_{DL,z}$ of 0.06 to match the write energy of the STT-MRAM and $\xi_{DL,z}$ of 0.13 to match the write energy of SRAM, the 2T-ZSOT with the half-fixed layer can approach the write energy of SRAM with a $\xi_{DL,z}$ of 0.051. Thus, the 2T-ZSOT half-fixed layer scheme can be a promising, dense, and energy-efficient option for future MRAM applications.

## References

[1] S. Yu, W. Shim, X. Peng and Y. Luo, "RRAM for Compute-in-Memory: From Inference to Training," in *IEEE Trans. Circuits and Systems I: Regular Papers*, vol. 68, no. 7, pp. 2753-2765, July 2021, doi: 10.1109/TCSI.2021.3072200.

[2] J. Okuno *et al*., "SoC Compatible 1T1C FeRAM Memory Array Based on Ferroelectric Hf0.5Zr0.5O2," *2020 IEEE Symposium on VLSI Technology*, Honolulu, HI, USA, 2020, pp. 1-2, doi: 10.1109/VLSITechnology18217.2020.9265063.

[3] D. Edelstein *et al*., "A 14 nm Embedded STT-MRAM CMOS Technology," in Proc. *2020 IEEE International Electron Devices Meeting (IEDM)*, San Francisco, CA, USA, 2020, pp. 11.5.1-11.5.4, doi: 10.1109/IEDM13553.2020.9371922.

[4] X. Han, X. Wang, C. Wan, G. Yu, and X. Lv, "Spin-orbit torques: Materials, physics, and devices," *Appl. Phys. Lett.*, vol. 118, no. 12, p. 120502, Mar. 2021, doi: 10.1063/5.0039147

[5] J. E. Hirsch, "Spin Hall Effect," *Phys. Rev. Lett.*, vol. 83, no. 9, pp. 1834–1837, Aug. 1999, doi: 10.1103/physrevlett.83.1834.

[6] D. Apalkov et al., "Spin-transfer torque magnetic random access memory (STT-MRAM)," *ACM J. Emerging Technologies in Computing Systems*, vol. 9, no. 2, pp. 1–35, May 2013, doi: 10.1145/2463585.2463589.

[7] T. Endoh, H. Honjo, K. Nishioka and S. Ikeda, "Recent Progresses in STT-MRAM and SOT-MRAM for Next Generation MRAM," in Proc. *2020 IEEE Symposium on VLSI Technology*, Honolulu, HI, USA, 2020, pp. 1-2, doi: 10.1109/VLSITechnology18217.2020.9265042.

[8] L. Liu et al., "Current-induced switching of perpendicularly magnetized magnetic layers using spin torque from the spin hall effect," Physical Review Letters, vol. 109, no. 9, p. 096602, Aug. 2012. doi:10.1103/physrevlett.109.096602

[9] M. Gupta et al., "High-density SOT-MRAM technology and design specifications for the embedded domain at 5nm node," in Proc. *2020 IEEE International Electron Devices Meeting (IEDM)*, San Francisco, CA, USA, 2020, pp. 24.5.1-24.5.4, doi: 10.1109/IEDM13553.2020.9372068.

[10] X. Li et al., "Materials Requirements of High-Speed and Low-Power Spin-Orbit-Torque Magnetic Random-Access Memory," *IEEE J. the Electron Devices Society*, vol. 8, pp. 674–680, Jan. 2020, doi: 10.1109/jeds.2020.2984610.

[11] W. Oh et al., "Field-free switching of perpendicular magnetization through spin–orbit torque in antiferromagnet/ferromagnet/oxide structures," *Nat. Nanotech.*, vol. 11, no. 10, pp. 878–884, Jul. 2016, doi: 10.1038/nnano.2016.109.

[12] C. Zhang, S. Fukami, H. Sato, F. Matsukura, and H. Ohno, "Spin-orbit torque induced magnetization switching in nano-scale TA/COFEB/MGO," *Applied Physics Letters*, vol. 107, no. 1, Jul. 2015. doi:10.1063/1.4926371.

[13] K. Garello *et al*., "Manufacturable 300mm platform solution for Field-Free Switching SOT-MRAM," *2019 Symposium on VLSI Circuits*, Kyoto, Japan, 2019, pp. T194-T195, doi: 10.23919/VLSIC.2019.8778100.

[14] M.-G. Kang, S. Lee, and B.-G. Park, "Field-free spin-orbit torques switching and its applications," npj Spintronics, vol. 3, no. 1, Mar. 2025, doi: https://doi.org/10.1038/s44306-025-00071-6.

[15] T. Zhao *et al*., "Enhancement of Out-of-Plane Spin–Orbit Torque by Interfacial Modification," *Advanced Materials*, vol. 35, no. 12, Jan. 2023, doi: https://doi.org/10.1002/adma.202208954.

[16] MacNeill, D. et al. (2016) 'Control of spin–orbit torques through crystal symmetry in WTE2/Ferromagnet bilayers', Nature Physics, 13(3), pp. 300–305. doi:10.1038/nphys3933.

[17] De Brosse, J. K., Liu, L., & Worledge, D. (2014). Spin hall effect assisted spin transfer torque magnetic random access memory, *U.S. Patent No. 8,896,041*. Washington, DC: U.S. Patent and Trademark Office.

[18] N. Sato, F. Xue, R. M. White, C. Bi, and S. X. Wang, "Two-terminal spin–orbit torque magnetoresistive random access memory," *Nat. Elec.*, vol. 1, no. 9, pp. 508–511, Sep. 2018, doi: 10.1038/s41928-018-0131-z.

[19] H. Zhang et al., "High-Density 1T1D1SOT-MRAM With Multimode Ultrahigh-Speed Magnetization Switching," *IEEE Magnetics Letters*, vol. 14, pp. 1–5, Jan. 2023, doi: 10.1109/lmag.2023.3293407.

[20] Y. -C. Liao *et al*., "Spin-Orbit-Torque Material Exploration for Maximum Array-Level Read/Write Performance," in Proc. *2020 IEEE International Electron Devices Meeting (IEDM)*, San Francisco, CA, USA, 2020, pp. 13.6.1-13.6.4, doi: 10.1109/IEDM13553.2020.9371979.

[21] S. Maekawa and T. Shinjo, "Spin Dependent Transport in Magnetic Nanostructures," *CRC Press*, 2002.

[22] Q. Hao and G. Xiao, "Giant Spin Hall Effect and Switching Induced by Spin-Transfer Torque in aW/Co40Fe40B20/MgOStructure with Perpendicular Magnetic Anisotropy," *Phys. Rev. Appl.*, vol. 3, no. 3, Mar. 2015, doi: 10.1103/physrevapplied.3.034009.

[23] L. Zhu, D. Ralph, and R. A. Buhrman, "Highly Efficient Spin-Current Generation by the Spin Hall Effect in Au1−xPtx," *Phys. Rev. appl.*, vol. 10, no. 3, Sep. 2018, doi: 10.1103/physrevapplied.10.031001.

[24] P. Kumar, Y.-C. Liao, D. C. Ralph, and Azad Naeemi, "A Drift-Diffusion Based Modeling and Optimization Framework for Nanoscale Spin-Orbit Torque Devices," *IEEE Trans. Electron Devices*, vol. 70, no. 2, pp. 789–795, Dec. 2022, doi: 10.1109/ted.2022.3227882.

[25] COMSOL, Inc., *COMSOL Multiphysics® Reference Manual*, Version 6.3, COMSOL, Inc., Burlington, MA, USA, 2023.

[26] M.N.H. Shazon, P. Kumar, and A. Naeemi, "Investigation on the impact of spin current profile on the write time of SOT MRAMs," In Proc. *Spintronics XVI*, San Diego, CA, USA, Aug. 2023, pp. 121–121, doi: https://doi.org/10.1117/12.2692161.

[27] Donahue, M. and Porter, D., "OOMMF User's Guide, Version 1.0."

[28] P. Kumar and A. Naeemi, "Benchmarking of spin–orbit torque vs spin-transfer torque devices," *Appl. Phys. Lett.*, vol. 121, no. 11, p. 112406, Sep. 2022, doi: https://doi.org/10.1063/5.010126.

[29] A. Meo *et al*, "Spin-transfer and spin-orbit torques in the Landau–Lifshitz–Gilbert equation," *J. Phys.: Cond. Mat.*, vol. 35, no. 2, pp. 025801–025801, Nov. 2022, doi: 10.1088/1361-648x/ac9c80

[30] L. Thomas, et al. "Perpendicular spin transfer torque magnetic random access memories with high spin torque efficiency and thermal stability for embedded applications." *Journal of Applied Physics,* vol. 115, pp.17615, 2014, doi: https://doi.org/10.1063/1.4870917.

[31] Sakhare *et al*., "Enablement of STT-MRAM as last level cache for the high performance computing domain at the 5nm node," *2018 IEEE International Electron Devices Meeting (IEDM)*, San Francisco, CA, USA, 2018, pp. 18.3.1-18.3.4, doi: 10.1109/IEDM.2018.8614637

[32] Bainsla *et al*., "Large out-of-plane spin–orbit torque in topological weyl semimetal Tairte4," *Nature Communications*, vol. 15, no. 1, May 2024. doi:10.1038/s41467-024-48872-3.